\documentstyle[epsfig]{aipproc2}
\begin{document}
\title{MAXIMA: an experiment to measure temperature\\ anisotropy
 in the cosmic microwave background}

\author{A.~T.\ Lee$^{1,2}$, P. Ade$^{3}$, A.  Balbi$^{1,4}$,
J. Bock$^{5,6}$, J. Borrill$^{1,7}$, A. Boscaleri$^{8}$,
B.~P. Crill$^{5}$, P. De~Bernardis$^{9}$, H. Del Castillo$^6$,
P. Ferreira$^{10}$, K. Ganga$^{11}$, S. Hanany$^{12}$,
V. Hristov$^{5}$, A.~H. Jaffe$^{1,2}$, A.~E. Lange$^{5}$,
P. Mauskopf$^{13}$, C.~B. Netterfield$^{14}$, S. Oh$^{1}$,
E. Pascale$^{5}$, B. Rabii$^{1,2}$, P.L. Richards$^{1,2}$,
J. Ruhl$^{15}$, G.~F. Smoot$^{1,2,4}$, C.~D. Winant$^{1,2}$}

\address{$^1$ Center for Particle Astrophysics,
  301 LeConte Hall, University of California, Berkeley, CA 94720\\
$^2$ Department of Physics, University of California, Berkeley, CA
94720\\
$^3$ Queen Mary and Westfield College, London, UK\\
$^4$ Lawrence Berkeley National Laboratory, Berkeley, CA 94720\\
$^5$ Department of Physics, California Institute of Technology, Pasadena, CA 91125\\
$^6$ Jet Propulsion Laboratory, Pasadena, CA 91125\\
$^7$ National Energy Research Scientific Computing Center, LBNL, Berkeley, CA
94720\\
$^8$ IROE-CNR, Firenze, Italy\\
$^9$ Dipartimento di Fisica, Universita La Sapienza, Roma, Italy\\
$^{10}$ CERN, Geneve 23, Switzerland and  
CENTRA, Instituto Superior Tecnico, Lisboa 1096 Codex, Portugal\\
$^{11}$ IPAC, Pasadena, CA 91125\\
$^{12}$ School of Physics and Astronomy, University of Minnesota, Minneapolis, MN 55455\\
$^{13}$ University of Massachusetts, Amherst, MA, 01003\\
$^{14}$ Department of Physics, University of Toronto, Toronto, Canada\\
$^{15}$ Department of Physics, University of California, Santa
Barbara, CA, 93106\\}

\maketitle

\begin{abstract}
We describe the MAXIMA experiment, a balloon-borne measurement
designed to map temperature anisotropy in the Cosmic Microwave
Background (CMB) from $\ell =80$ to $\ell =800$.  The experiment
consists of a 1.3~m diameter off-axis Gregorian telescope and a
receiver with a 16 element array of bolometers cooled to 100 mK. The
frequency bands are centered at 150, 240, and 410 GHz.  The
10$^\prime$ FWHM beam sizes are well matched to the scale of acoustic
peaks expected in the angular power spectrum of the CMB.  The first
flight of the experiment in its full configuration was launched in
August 1998.  A 122 deg$^2$ map of the sky was made near the Draco
constellation during the 7 hour flight in a region of extremely low
galactic dust contamination.  This map covers 0.3\% of the sky and has
3200 independent beamsize pixels.  We describe the MAXIMA instrument
and its performance during the recent flight.
\end{abstract}

\section*{Introduction}
%
%
%
%

Observations of the cosmic microwave background (CMB) temperature
anisotropy are providing strong constraints on theories of
cosmological structure formation.  These measurements discriminate
between competing cosmological models and, if the standard
inflationary model is correct, will determine each many of the
fundamental cosmological parameters to high precision.

In this paper we describe the MAXIMA experiment, a balloon-borne
telescope designed to measure the angular power spectrum of CMB
anisotropy over a wide range of angular scales.  The experiment
consists of a 1.3 m diameter off-axis Gregorian telescope and a
receiver with a 16 element array of bolometers cooled to 100 mK.  The
high sensitivity of this receiver allows accurate measurements of the
CMB power spectrum in a single overnight balloon flight.  Each of the
pixels in the array are sensitive to a single frequency band centered
at 150, 240, or 410 GHz.  The 150 GHz band is the most sensitive to
the CMB and is close in frequency to the predicted minimum in galactic
foregrounds. The higher frequency channels monitor emission from the
atmosphere and galactic foregrounds such as dust.  The combination of
a small beam size (10$^\prime$ FWHM) and long scans (4$^\circ$ p-p)
result in sensitivity from angular scales of 4$^\circ$ to 10$^\prime$
($80 < \ell < 800$, where $\ell$ is the spherical harmonic multipole
number.) This range is well matched to the expected acoustic peaks in
the CMB angular power spectrum.

The scan strategy for MAXIMA has been carefully designed to be
efficient and to minimize and measure systematic effects.  The
observation is done in a total power rather than chopped mode. The sky
is scanned with three modulations: a primary mirror scan, a gondola
scan, and the same patch of sky is scanned twice during the flight.
These modulations allow interference on any one time scale to be
isolated.  The two scans of the sky are performed at different angles
due to sky rotation to ``cross link'' the scans and reduce striping
due to 1/f noise in the detector.  This cross-linking results in a
more sensitive measurement of the angular power spectrum.

The first science flight MAXIMA-1 was launched in August 1998.  All
systems worked well and bolometer loading and sensitivity were
consistent with those estimated with ground-based measurements.  A 122
deg$^2$ map of the sky was made near the Draco constellation during
the 7 hour flight in a region of extremely low galactic dust
contamination.  This map covers 0.3\% of the sky and has 3200
independent beamsize pixels.

MAXIMA is part of the MAXIMA/BOOMERanG collaboration (see papers by
P. De Bernardis and S. Masi in these proceedings).  MAXIMA and
BOOMERanG share a common base of technologies, including bolometric
detectors, optical filters, and attitude control systems.  MAXIMA is
optimized for multiple single-day flights whereas BOOMERanG is
designed for a long duration flight around the antarctic continent.

MAXIMA is a valuable testbed for the technologies planned for the
High-Frequency Instrument (HFI) of the Planck Explorer.  The 100 mK
``spiderweb'' bolometers built at JPL are similar to those planned for
the HFI.\cite{bock98} MAXIMA has a similar beamsize to the HFI at 150
GHz and also uses single-color feedhorns in the focal plane.

In the following sections, we will describe the optical system,
receiver design, detector system, gondola, attitude control system,
scan strategy, calibration strategy, and performance during the recent
flight.

\label{sec:methodsLike}

\section{Optical system}
The MAXIMA telescope is an underfilled f/1 off-axis Gregorian design
with two cold reimaging mirrors as shown in Fig. 1.  The primary
mirror is a 1.3~m diameter off-axis section of a paraboloid.  The
secondary and tertiary mirrors (21 cm and 18 cm in diameter
respectively) are off-axis ellipsoids with aspheric components
which compensate the part of the aberrations introduced by the primary
mirror.  This optical system provides a 1$^\circ$ x 1$^\circ$
diffraction-limited field-of-view at 150 GHz.  A baffle between the
two reimaging mirrors and a Lyot stop provide excellent telescope
sidelobe performance. The overall design has many elements in common
with the DIRBE telescope for the same reasons.  The reimaging optics
and all baffles are maintained at LHe temperature to reduce the
optical power loading on the bolometers, thereby improving
sensitivity.  The non-optical interior surfaces of the box are covered
with a mm-wave absorbing material \cite{bock97}.  A 1\%
neutral-density filter can be inserted in the optical path at the
intermediate focus for tests using 300K loads. The neutral-density
filter is inserted and removed using a bellows-sealed screw on the
outside of the cryostat.

\begin{figure}[t!] 
\centerline{\epsfig{file=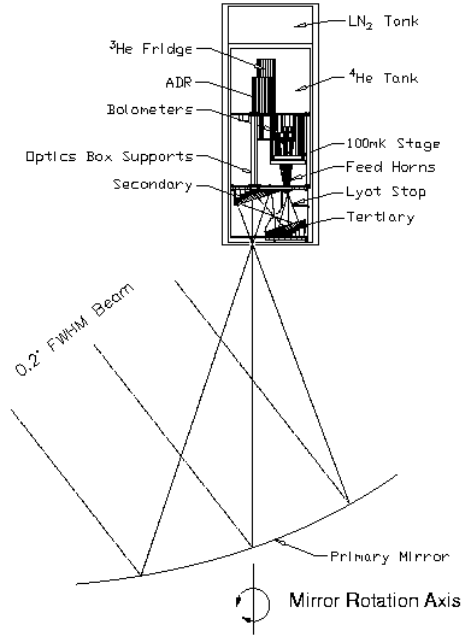,height=3.5in, angle=0}}
\caption{Cross-section of optical system.  The primary mirror is a 1.3
m diameter off-axis paraboloid.  The two reimaging mirrors are housed
in a well baffled box that is maintained at LHe temperature.  The
optical system provides a 1$^\circ$ x 1$^\circ$ diffraction-limited
field-of-view at 150 GHz. A baffle at the intermediate focus and a
Lyot stop provide excellent telescope sidelobe performance.  The
bolometers are cooled to 100 mK by an Adiabatic Demagnetization
Refrigerator.  Both LN and LHe cryogen hold times are $\approx40$
hours.  The optical entrance to the receiver is vacuum sealed with a
window made from 40 $\micron$ thick polypropylene.}
\vspace*{10pt}
\label{fig1}
\end{figure}

The focal plane contains 16 single-color feedhorns designed for
10$^\prime$ FWHM beamsize as shown in Figs. 2-4.  The focal plane is
organized into four equal elevation rows each containing two 150 GHz
feeds, a 240 GHz feed, and a 410 GHz feed.  Two types of feedhorn are
used in the focal plane.  For the 150 GHz channels conical horns are
used, since these channels are at the diffraction limit.  The conical
horns are smooth walled for ease of fabrication. The sidelobe
performance of the horn is not critical given the use of a Lyot stop.
For the 240 and 410 GHz channels, Winston horns are used since these
channels detect multiple optical modes.

\begin{figure}[t!] 
\centerline{\epsfig{file=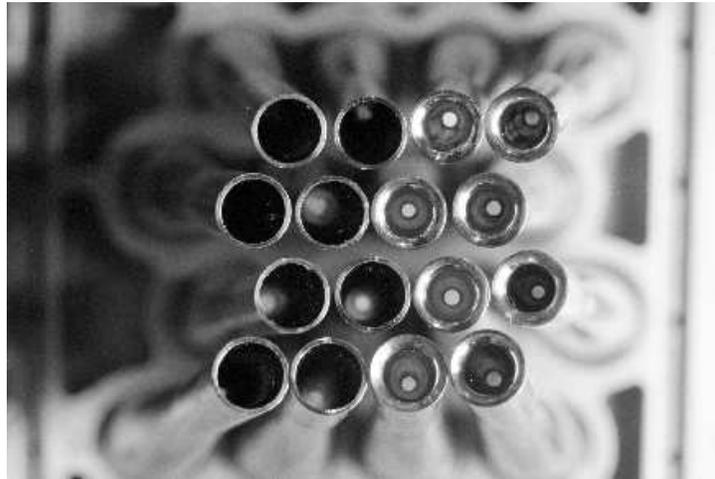,height=2.5in,angle=0}}
\caption{Photograph of MAXIMA focal plane.  Smooth walled conical
horns are used for the eight diffraction-limited 150 GHz channels (two
columns to left), and Winston horns are used for the 240 and 410 GHz
channels (two columns to right) which detect multiple optical modes.
As the array is scanned in azimuth (horizontal in photo), each pixel
is measured with four detectors in a row with the three frequency
bands.  The beamsize for all pixels is 10$^\prime$. The outer diameter
of the horns is $\approx 5.8$ mm.}
\vspace*{10pt}
\label{fig2}
\end{figure}

\begin{figure}[t!] 
\centerline{\epsfig{file=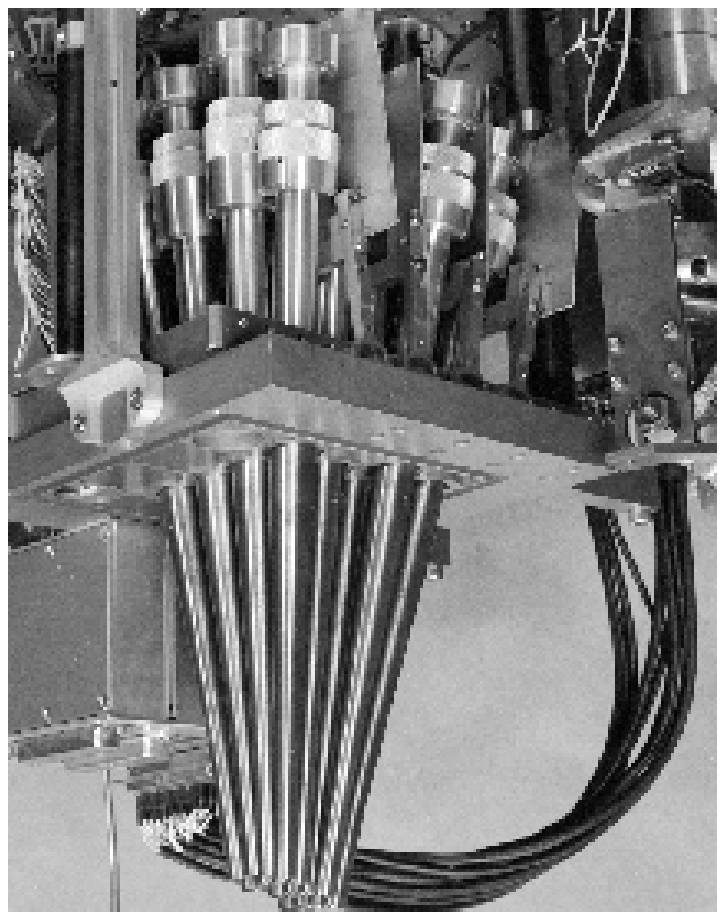,height=3.5in, angle=180}}
\caption{Photograph of the interior of the MAXIMA receiver.  The focal
plane is at the bottom with horns and light pipes extending to the LHe
temperature mounting plate.  The 100 mK plate which carries the
metal-mesh filter holders and bolometer integrating cavities is spaced
0.5 mm above the LHe plate.  The $^3He$ refrigerator and heat switch
used to conduct the ADR heat of magnetization to $^3He$ refrigerator
are visible on the upper right.  Each aluminum box (bottom left)
contains FET front-end electronics for five channels.}
\vspace*{10pt}
\label{fig3}
\end{figure}

The frequency bands are defined by metal-mesh filters built by the
QMWC group.  These are positioned inside light pipes between the
feedhorns and the bolometers as shown in Fig. 4.  For the 150 GHz
channels a high-pass filter is formed by a small circular waveguide at
the back of the feedhorn.  Spectra measured before the flight are
shown in Fig. 5.  High-frequency out-of-band leaks are dangerous due
to the rapidly rising spectrum of the atmosphere.  Three low-pass
filters (two metal mesh and one made from alkali-halide powder in
polyethylene) are used to block resonant leaks in the band-defining
filters and to reduce optical loading on the cryogens. We tested for
leaks above the passband for each channel by comparing bolometer
response with and without a thick-grill high-pass filter blocking a
chopped source.  The cutoff frequency of the thick-grill filter was
chosen to be just above the upper edge of the passband.  Leak
performance was found to be greatly improved in channels with two
metal mesh filters (see Fig. 4) by introducing an absorbing section of
lightpipe made from G-10 in between the two filters.  The inner
diameter of this lightpipe was chosen to be the same as that of the
adjacent lightpipe.  This technique, which may suppress multiple
reflections between the filters, did not adversely affect in-band
efficiency.

The feedhorns are supported by an aluminum plate at LHe temperature.
Another plate, at 100 mK, supports the filter and bolometer
assemblies.  The two plates are separated by a 0.5 mm gap.  The 100 mK
side of the gap is covered with absorber to prevent stray light from
entering \cite{bock97}.

The bolometers are housed in integrating cavities fed by conical
horns.  Several different backshort geometries were evaluated including a
flat plate $1/4 \lambda$ from the bolometer, a spherical cavity, and a
sloping flat backshort.  System optical efficiencies of most channels
were 25-35\%, with no significant differences between the
different backshort geometries.

\begin{figure}[t!] 
\centerline{\epsfig{file=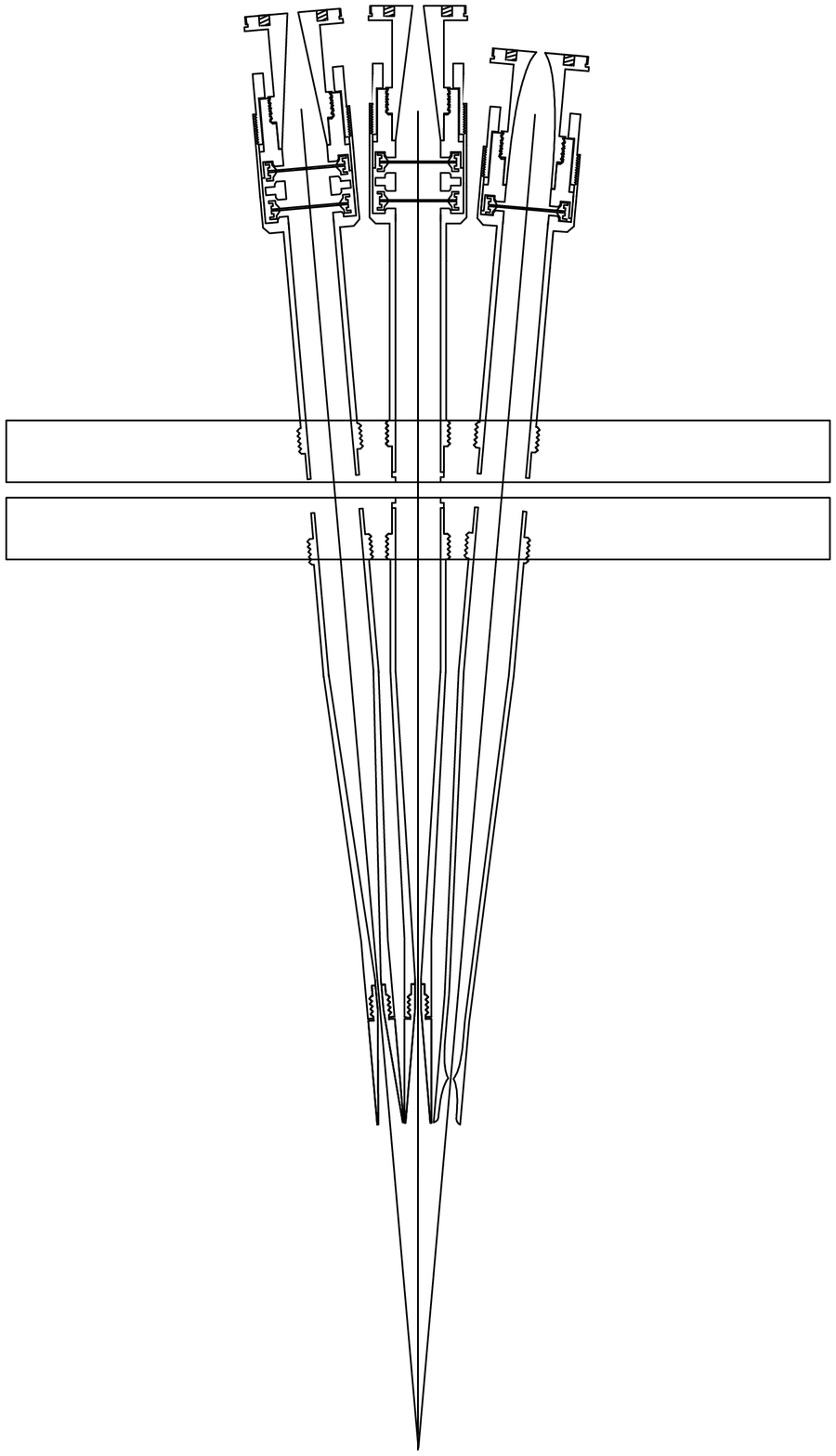,height=3in, angle=180}}
\caption{Cross-section view of the detector array.  The single mode
150 GHz channels (two horns on left) use smooth-wall conical feedhorns
since they are diffraction limited to one optical mode.  The 240 and
410 GHz channels (only one horn shown on right) use Winston horns
since they are sensitive to multiple optical modes. The frequency
bands are defined by metal-mesh filters positioned within light pipes
between the feedhorns and the bolometers.  These filters are built by
the QMWC group. For the 150 GHz channels a high-pass filter is formed
by a small circular waveguide at the back of the feedhorn. The
feedhorns are mounted to a plate maintained at LHe temperature.  The
filters and bolometers are mounted to a 100 mK plate that is separated
by 0.5 mm from the LHe temperature plate.  The bolometers and
integrating cavities (not shown) are mounted behind the horns seen at
the top of the drawing.}
\vspace*{10pt}
\label{fig4}
\end{figure}

\begin{figure}[t!] 
\centerline{\epsfig{file=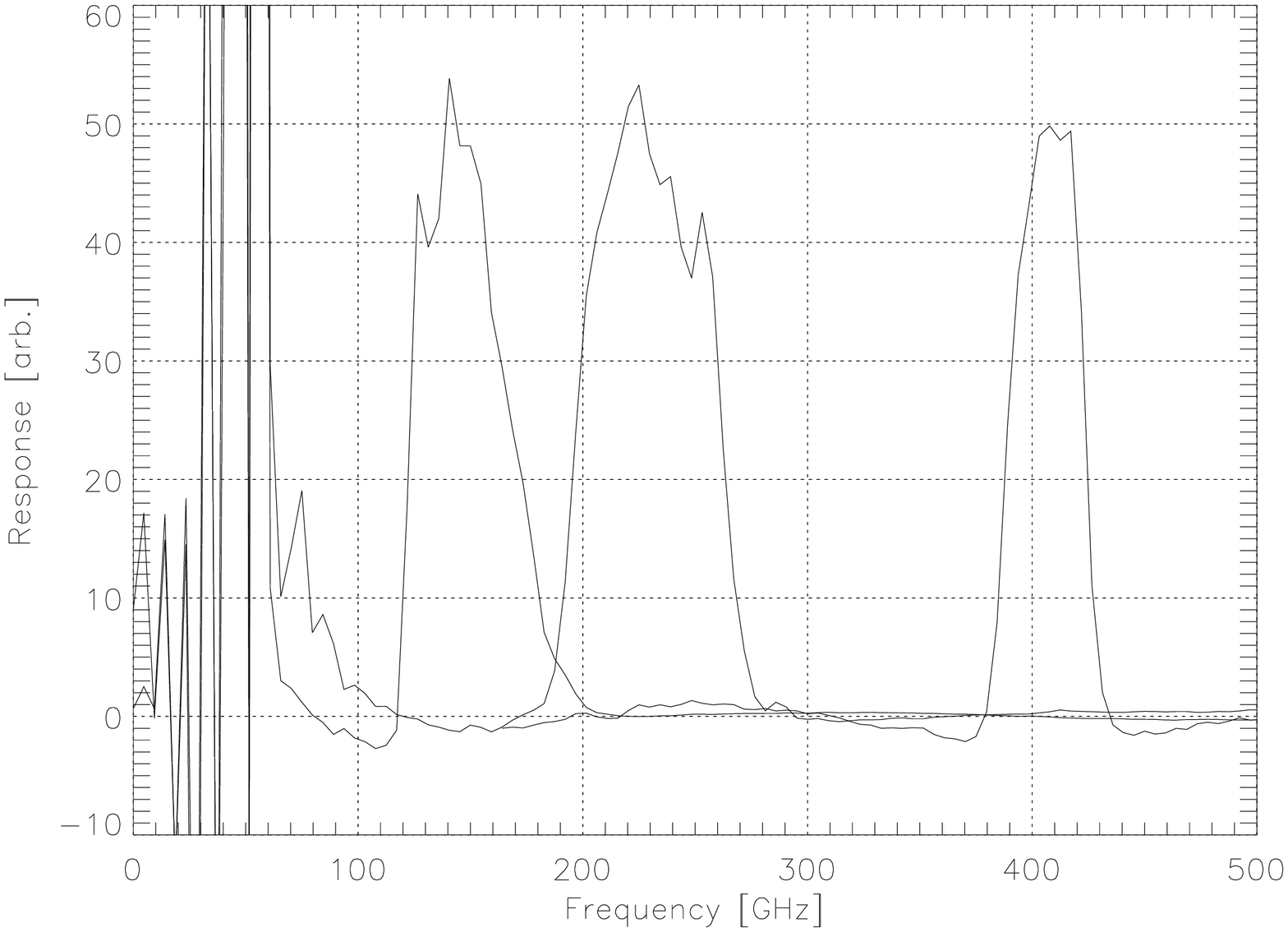,height=3in, angle=90}}
\caption{Spectral response of detector system.  The data were obtained
with a far-infrared Fourier spectrometer.  The receiver was in flight
configuration.  The high noise below 100 GHz results from the measurement
method and does not reflect detector response.}
\vspace*{10pt}
\label{fig5}
\end{figure}

The beams are scanned on the sky by rotating the primary mirror about
an axis defined by the center of the primary mirror and the prime
focus.  We use a lightweight (11 kg) mirror constructed with an
aluminum honeycomb core, graphite facesheets, and a sputtered aluminum
optical surface.  The mirror is rotated by a large dc servo motor.
The position is sensed by a Linear-variable-differential transformer
(LVDT), and the position is controlled by an analog
proportional-differential feedback system.  The reference scan is a
0.45 Hz 4$^\circ$ p-p sawtooth with smoothed transitions.  Data from
the flight is shown in Fig. 6.

\begin{figure}[t!] 
\centerline{\epsfig{file=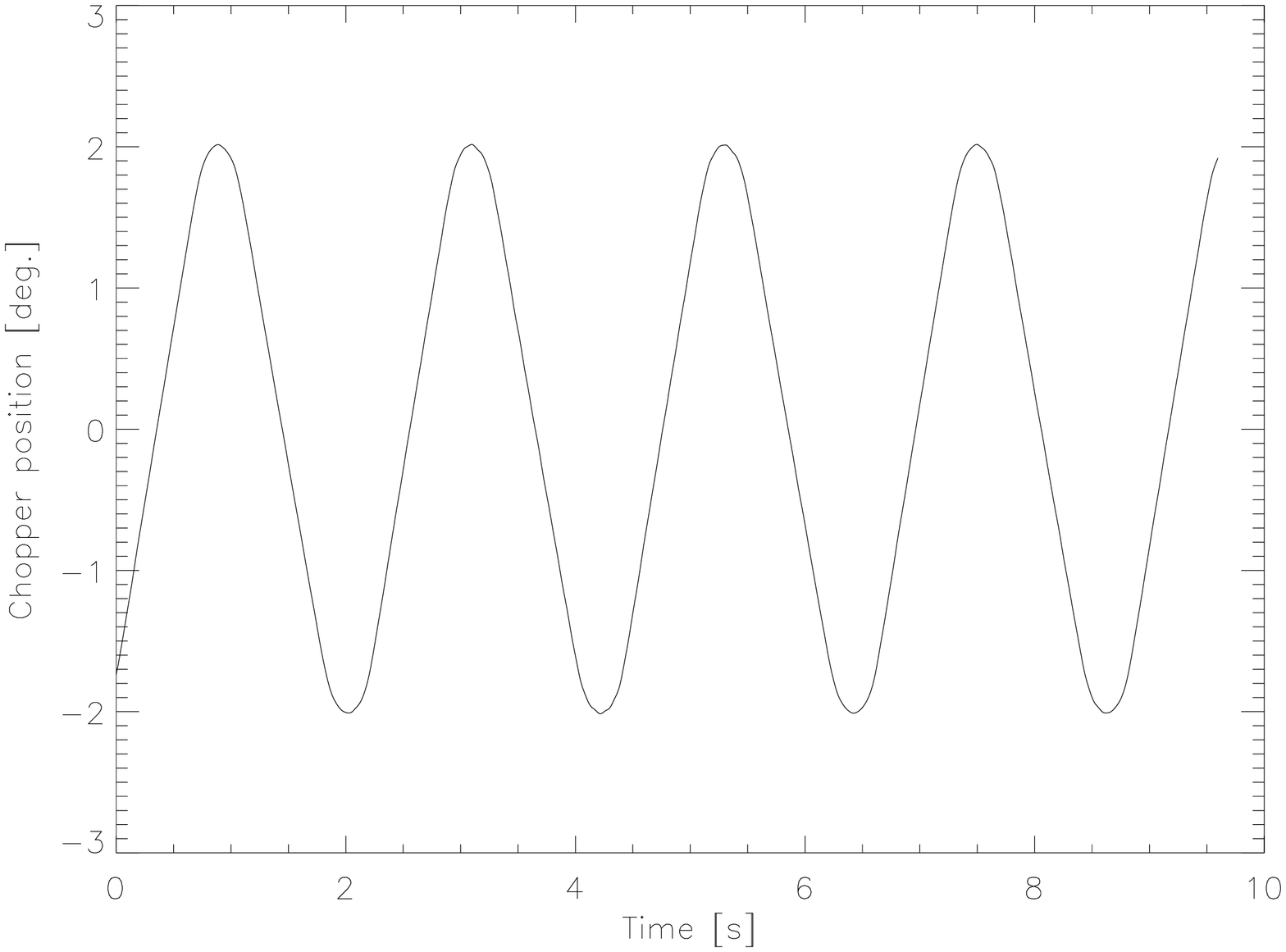,height=3in, angle=90}}
\caption{Measured azimuthal position of primary mirror during flight.
The frequency is 0.45 Hz and the p-p scan is $4^\circ$ on the sky.}
\vspace*{10pt}
\label{fig6}
\end{figure}

Beam patterns were measured using Jupiter as a source during the
MAXIMA-1 flight.  Single azimuthal scans are shown in Fig. 7. for four
channels.  The beam sizes are consistent with the design goal of
10$^\prime$ FWHM.  Scans were performed at multiple elevations to
measure the beam patterns in two dimensions.

\begin{figure}[t!] 
\centerline{\epsfig{file=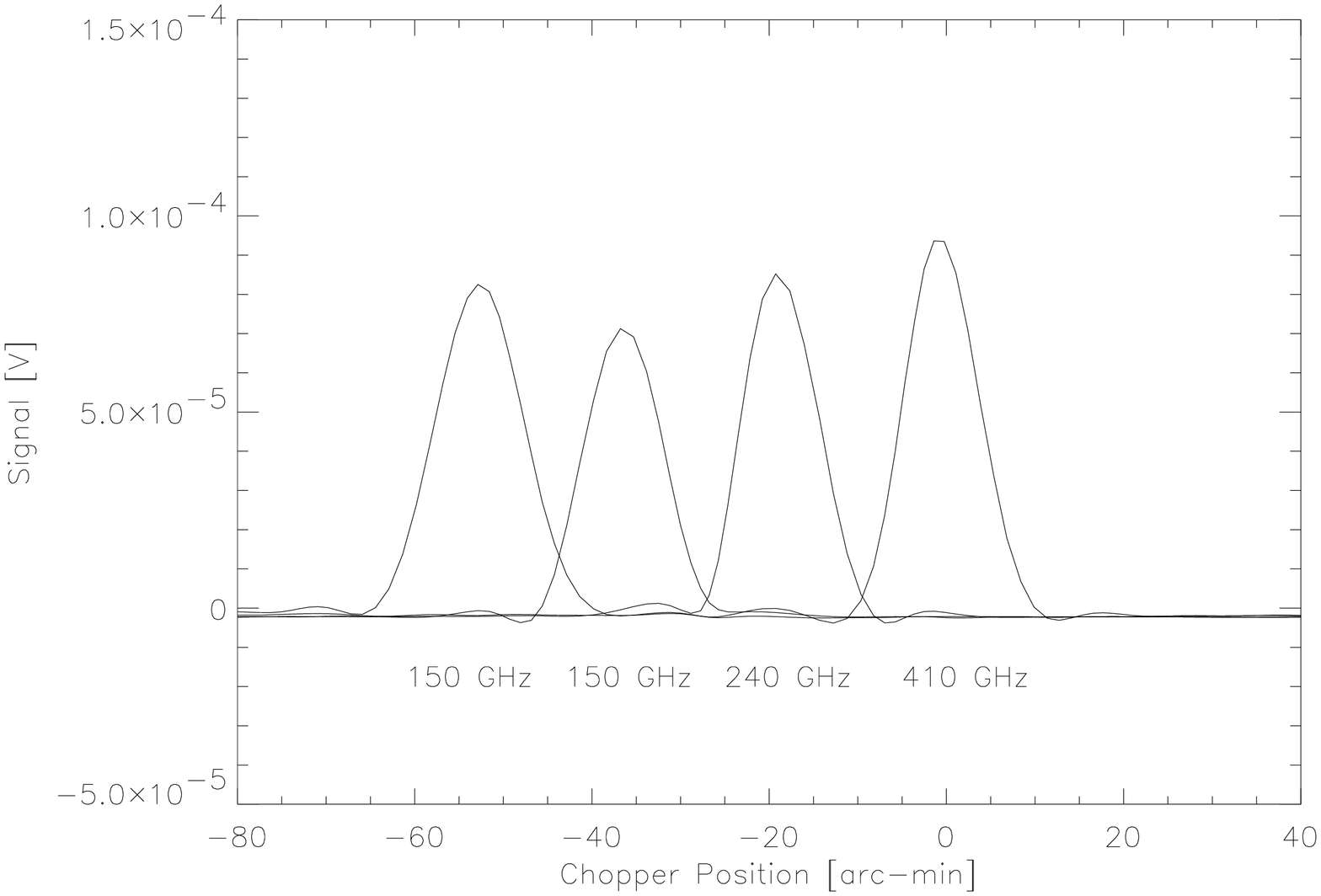,height=3in, angle=90}}
\caption{Beam patterns of one row of detectors using Jupiter as a
source.  Signals represent a single scan of the mirror in
azimuth. Beam widths are 10-12$^\prime$ FWHM, without taking into account the
finite size of Jupiter.  There is some slight ringing in the baselines
due to a combination of electronic and bolometer time constants.
Detector noise is not visible on this plot due to the high
signal-to-noise ratio.}
\vspace*{10pt}
\label{fig7}
\end{figure}

The sidelobe performance was measured on the ground for the 150 GHz
band as shown in Fig. 8.  A 50 mW 150 GHz Gunn oscillator source was
placed on the roof of the NSBF highbay (height $\approx 37$~m) with the
gondola in the parking lot in flight configuration.  One telescope
beam was pointed at the source and then the telescope was rotated
first in elevation and then in azimuth to measure the sidelobe
response.  A variable attenuator was used on the source to keep the
signal in the linear regime of the bolometers.  In addition to the
scans in elevation and azimuth the source was directed at all parts of
the gondola from ground level.

\begin{figure}[t!] 
\centerline{\epsfig{file=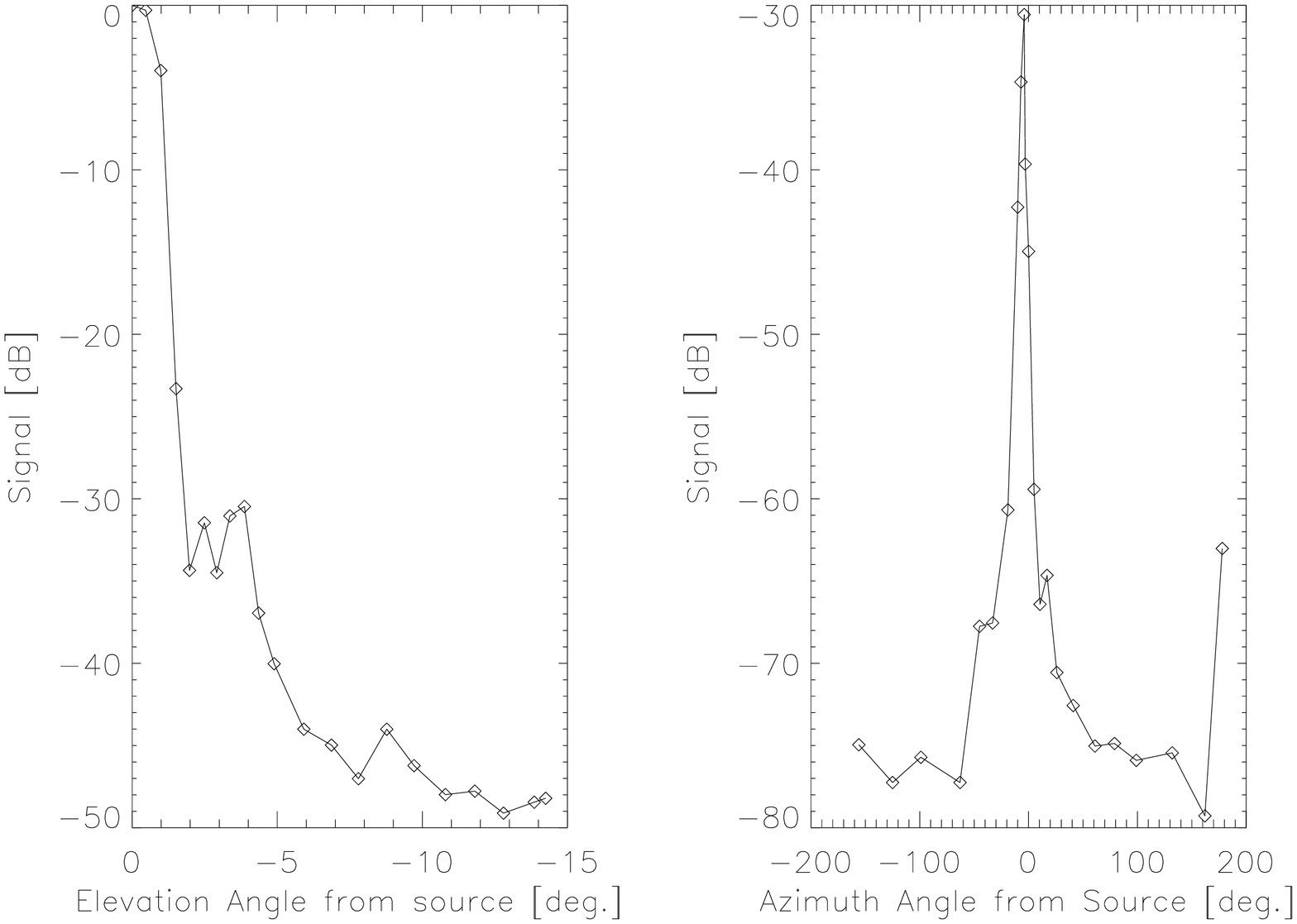,height=6in,width=3in,angle=90}}
\caption{Sidelobe patterns for the MAXIMA telescope at 150 GHz.  The
near sidelobe pattern in elevation is at the left, and far sidelobe
pattern in azimuth is at the right.  Measurements were done with
source on top of the NSBF highbay with a source elevation of
$42^\circ$.  Adjustable attenuators were used on the source to
keep the detectors in the linear regime.  The azimuth measurements
were done with the elevation $\approx 2^\circ$ below the elevation of
the source.  This suppressed the main lobe during azimuth scan.}
\vspace*{10pt}
\label{fig8}
\end{figure}

\section{Receiver Design}

The MAXIMA cryostat houses the secondary optics, the bolometers, and
the sub-Kelvin coolers.\cite{irlabs} These are all mounted to a ``cold
plate'' which forms one side of the LHe tank, and are surrounded by
radiation shields at LHe and LN temperatures.  The liquid nitrogen
(liquid helium) tanks have a 13 (21) liter capacity, and the unpumped
hold times are $\approx40$ hours for both cryogens.

The bolometers are cooled to 100 mK by an Adiabatic Demagnetization
Refrigerator.  A detailed description of this device can be found in
Hagmann {\it et al.}\cite{hagmann} The salt pill is made from Ferric
Ammonium Alum (FAA).  The superconducting magnet uses a NbTi coil
and generates 2.5~T for a peak current of 6.2~A.  The magnet current is
carried from the LN temperature stage to the LHe temperature stage by
commercial leads made from High-$T_c$ superconductor, which give a
heat leak of only $\approx 5$ mW.\cite{hightc} The use of these leads
gives a 15\% decrease in total heat leak to the LHe compared to using
normal metal leads.  The cycle time of the ADR is 2 hours, and the
hold time at 100 mK is $\approx48$ hours. The heat of magnetization is
absorbed by a 300 mK closed-cycle single-shot charcoal-pumped $^3He$
refrigerator, which also provides a heat intercept to the 100 mK
stage.  The refrigerator contains 40 STP liters of $^3He$ and uses an
external tank to reduce the internal pressure when the system is at
room temperature.  The refrigerator takes 4 hours to cycle and has a
hold time of 40 hours.

The heat leak to the 100 mK stage has been minimized through the
choice of materials and geometry of the supports and wires.  The
bolometer platform is mounted to the liquid helium cold plate with
three tubes (0.84 cm diameter, 17.8 cm long, and 0.5 mm wall
thickness) made from Vespel SP-22 plastic which is strong but gives
low heat leak at sub-Kelvin temperatures.  The tubes are heat sunk to
300 mK at 1/3 of their length from the LHe temperature end.  The 120
Pt-W bolometer wires are 5 cm long and have a 51 micron diameter.  The
signal wires are twisted with a ground wire and are attached to G-10
tubes (wrapped and varnished) to reduce microphonic response.  The
G-10 tubes are 10.2 cm long, 0.5 cm in diameter, have 130 micron wall
thickness, and the heat flow is intercepted in a similar manner to the
Vespel support legs.

We use a modular ``wire insert'' to carry all wires from the top of
the cryostat at 300 K to the cold plate at LHe temperature.  This
insert, shown in Fig. 9, provides easy access to the wires.  It is
inserted through holes in the cryostat shell and the LN and the LHe
tanks and is then mechanically attached at each temperature stage.  It
contains the 300 K connectors and essentially all of the cryostat
wiring. It also includes cold RF filter modules made from striplines
encased in lossy Eccosorb CR-124.

\begin{figure}[t!] 
\centerline{\epsfig{file=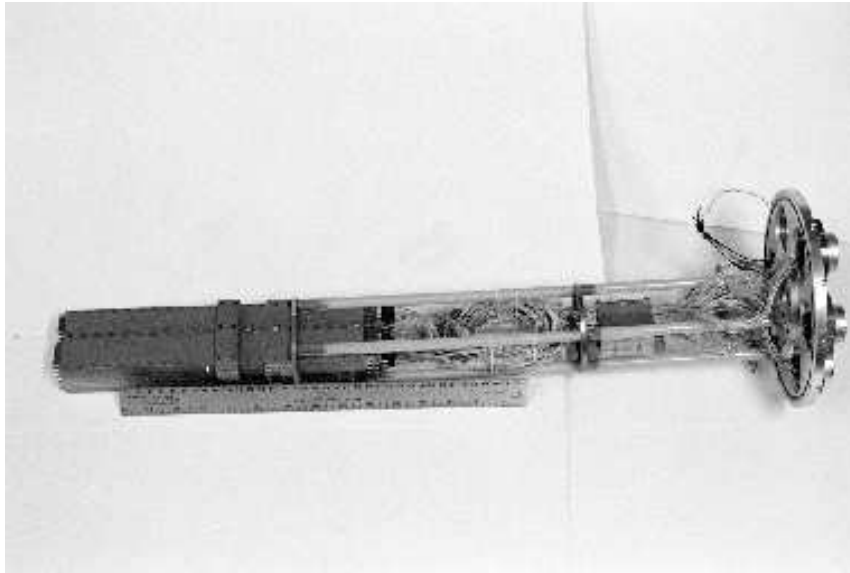,height=3in,angle=0}}
\caption{Photograph of wire insert.  The modular wire insert is
introduced through holes in the cryostat shell and LN and LHe tanks,
and then mechanically attached to each temperature stage.  The wire
insert contains the 300 K connectors, essentially all of the cryostat
wiring, and cold RF filter modules which consist of striplines encased
in lossy Eccosorb CR-124.  The three- and five-conductor stainless
steel wires with braided stainless shields are organized into five
wiring modules, one each for five bolometer channels, and one for
housekeeping wires.  Each of these modules can be easily replaced if
found to be faulty.}
\vspace*{10pt}
\label{fig9}
\end{figure}

\section{Detector System}

We use spiderweb-absorber bolometers fabricated at JPL with
Neutron-Transmutation Doped (NTD) Ge thermistors, which give high
sensitivity with a fast response time, low cosmic-ray cross section,
and low microphonic response.\cite{bock98} The NTD Germanium material
is produced at LBNL.  Time constants of $\tau\approx10$ ms have been
measured which are essential for our rapid primary mirror modulation.
For our total power mode scan, the speed at which the beam can be
modulated on the sky is limited to $\approx FWHM/2\tau$, where FWHM
refers to the full-width half-maximum beamsize.\cite{hanany98}

The bolometer signals are buffered using differential JFET amplifiers.
We use commercial devices from Infrared Laboratories that are
internally thermally isolated with polyamide legs.\cite{irlabs} These
devices dissipate $\approx 1$ mW and have a white noise level of
$\approx 9~nV/\sqrt Hz$.

The bolometers are ac-biased at 220 Hz using a sinusoidal bias and the
signals are band-pass filtered and then lock-in detected. This ac-bias
scheme circumvents 1/f noise in the front-end amplification and
results in excellent low-frequency
stability.\cite{wilbanks90,devlin93} A noise spectral density from the
flight is shown in Fig. 10.  The noise below 0.1 Hz is largely
attributable to temperature fluctuations in the unregulated 100 mK
stage.

A 4.2K temperature optical load was used in ground tests to check the
optical loading without the neutral-density filter.  The optical
loading on the bolometers in flight was consistent with that predicted
from these measurements. During the MAXIMA-1 flight four of the eight
150 GHz channels had a Noise Equivalent Temperature (NET-CMB) of
90-100 $\mu K\sqrt sec$.  Two others had NET $\approx150 ~\mu K\sqrt
sec$, and the last two had significantly higher noise.  The
variability in sensitivity reflects a variation in noise more than in
responsivity.  Some excess noise came from the ac-bias readout system.
The sensitivity of all 150 GHz channels taken together in quadrature,
which is the effective receiver sensitivity at 150 GHz, was 41 $\mu
K\sqrt sec$.

We employ an optical ``stimulator'' located adjacent to the Lyot stop
to provide a periodic calibration reference for the bolometers.  This
stimulator is similar to a bolometer in construction and uses a
resistive heater to heat it to several tens of degrees Kelvin.  It is
turned on for 10 seconds every 25 minutes.  We use three monitor
channels in addition to the 16 optical bolometers.  These include an
optically blanked bolometer and a 4 MOhm resistor, which are used to
monitor electro-magnetic interference.  We also use a well heat sunk
NTD thermistor to monitor temperature fluctuations of the 100 mK
stage.

\begin{figure}[t!] 
\centerline{\epsfig{file=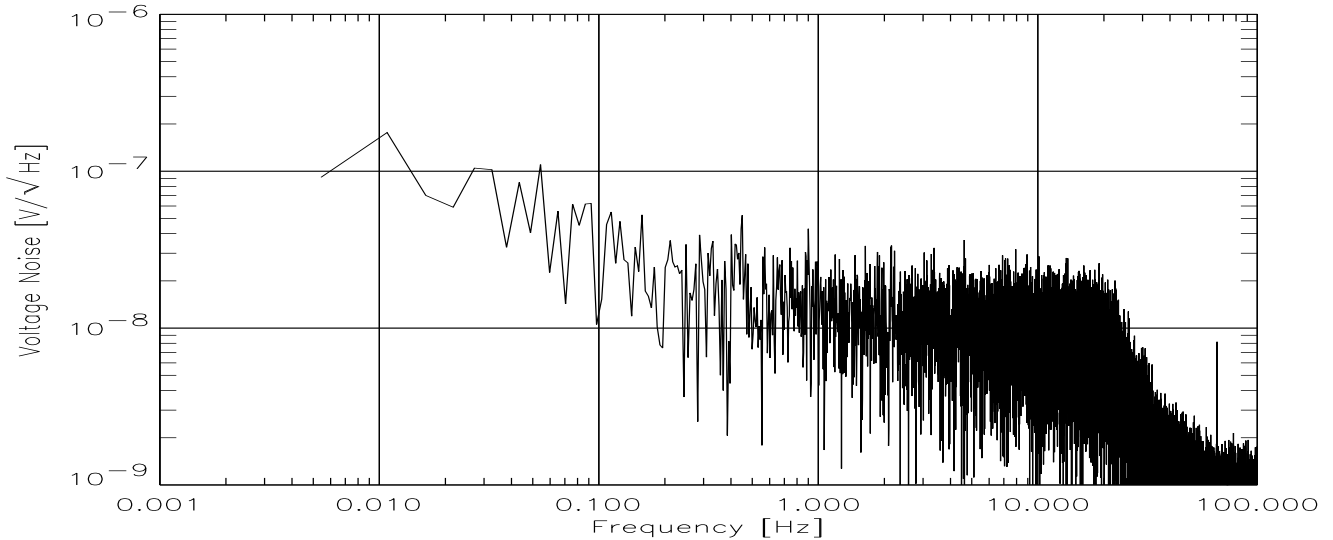,height=7in,width=3in,angle=90}}
\caption{Noise referred to bolometer for 150 GHz channel during CMB
scan. The bolometers are ac-biased at 220 Hz using a sinusoidal bias
and the signals are band-pass filtered and then lock-in detected.  The
effect of the 20 Hz low-pass filter in the readout electronics can be
seen to the right.  Low-frequency noise is largely attributable to
temperature fluctuations in the unregulated 100 mK stage.}
\vspace*{10pt}
\label{fig10}
\end{figure}

\section{Gondola and attitude control system}

The gondola, shown in Fig. 11, is designed to be strong enough for
multiple flights. It incurred little damage in the MAXIMA-0 test
flight in 1995 or in MAXIMA-1.  The design is based on the ACME
gondola used with the MAX experiment and is largely constructed with
bolted aluminum.  The gondola has a sturdy outer frame made from
members with an ``L'' shaped cross-section, with tubular roll bars
mounted at the corners.  The receiver and primary mirror are mounted
on an inner frame supported on trunion bearings to change the beam
elevation.  The elevation range is $20^\circ$ to $55^\circ$.  The
attitude control electronics, data acquisition, and command
electronics are housed in boxes with aluminum honeycomb sides on the
either side of the gondola.  The bottom of the gondola is covered in
aluminum honeycomb.  The flight ready gondola weights 3600 lbs without
ballast.

\begin{figure}[t!] 
\centerline{\epsfig{file=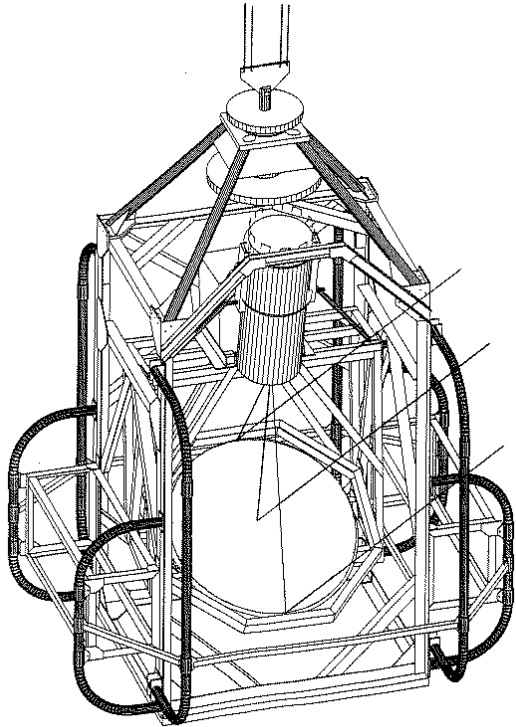,height=4in,angle=0}}
\caption{Drawing of the MAXIMA gondola. The primary mirror and the
receiver are mounted to an inner frame supported on trunion bearings
to change beam elevation from $20^\circ$ to $55^\circ$.  The gondola
is largely constructed with bolted aluminum members.  The attitude
control and data acquisition electronics are housed in the two
aluminum boxes on the sides of the gondola.}
\vspace*{10pt}
\label{fig11}
\end{figure}

The attitude control system (ACS) controls the position of the
telescope beams on the sky.  The ACS consists of attitude sensors, a
central feedback loop computer, and actuator motors.  The ACS can be
divided into one system that controls the azimuthal angle of the
gondola and another that controls the elevation of the inner frame.
For both systems, only some attitude sensors are used in the actuator
control loop, while all are used for post-flight pointing
reconstruction.

The azimuthal control portion of the ACS employs a pivot unit which
contains two dc motors.  One motor drives a reaction wheel and the
other drives directly against the flight line.  All scans are done
with constant gondola azimuthal velocity.  The feedback for the
reaction wheel is proportional to azimuthal velocity as measured by an
azimuthal rate gyroscope.  The feedback for the direct drive motor is
proportional to the reaction wheel velocity as measured by a
tachometer.  This loop prevents the reaction wheel velocity from
reaching saturation.  

For the CMB anisotropy scan, the velocity is reversed periodically to
produce a sawtooth scan.  A two-axis magnetometer is used to determine
the turnaround locations.  During the MAXIMA-1 flight azimuthal
velocity error was typically $\approx 0.3^\prime/s$ with velocities during
the anisotropy scans of $0.3-0.5^\circ/s$.  This performance is more
than adequate since small errors in gondola velocity only affect the
uniformity of sky coverage.

The telescope elevation is changed using a ballscrew and a small dc
servo motor.  The position is sensed using a 16-bit optical encoder. A
digital feedback loop similar to the azimuth loop is used for control.
The telescope elevation is not actively controlled during CMB scans.
The telescope has a range in elevation of 20$^\circ$ to 55$^\circ$.

For post-flight pointing reconstruction of the CMB scans, the azimuth
and elevation of the beam are detected with two CCD cameras.  The
boresight camera has a 7$^\circ$ field of view in azimuth, and can
lock on stars as faint as 6th magnitude.  The other camera is attached
to the outer frame, has a 14$^\circ$ field of view, and is offset in
azimuth to lock on Polaris during the CMB scans.  The two-axis
magnetometer is used for pointing reconstruction during azimuthal
rotations which were used to calibrate from the CMB dipole.  Roll and
pitch rate gyroscopes are also used in the pointing reconstruction.

Gondola pendulations give signals in the bolometers due to the change
in optical path length through the atmosphere with beam elevation.
The most troublesome mode observed during the 1995 MAXIMA-0 test
flight was the one in which the gondola rotates about its center of
mass.\cite{msam} For our gondola, the frequency of this mode is 0.6
Hz, which is in our signal bandwidth.  The lower frequency modes are
not important during the CMB scans, since they occur at frequencies
much lower than our 0.45 Hz mirror scan.  Four approaches were used to
reduce the 0.6 Hz pendulation.  First, the reaction wheel was made as
symmetric as possible by using well measured weights at the end of
lightweight spokes.  Second, the axis of the reaction wheel was
aligned to the gravity vector to an accuracy of $\approx1^\prime$
before flight.  Third, the azimuth feedback loop uses a sensor
(gyroscope) that has no sensitivity to pitch or roll.  Finally, we use
a passive pendulation damper built by Geneva Observatory, which
consists of a damped harmonic oscillator whose resonance frequency is
set to the 0.6 Hz pendulation frequency.\cite{genevaobs} This device is
constructed from a spherical weight rolling on a spherical cup in an
oil filled vessel.  We calculate that this damper should reduce
pendulation amplitude by a factor of $\approx 10$.  During the
MAXIMA-1 flight the amplitude of the 0.6 Hz pendulation mode was an
acceptable $1-10^{\prime\prime}$ rms.

\section{flight}

The MAXIMA-1 science flight was launched on August 2, 1998 at 00:58 UT
(7:58 PM local) from the NSBF facility at Palestine Texas (see
Fig. 12.).  We reached a float altitude of 37.5 km at 4:35 UT and
remained above 36.6 km for 3 hours and 47 minutes until the flight was
terminated.  The time at float was somewhat short due to the fast
high-altitude winds during mid-summer.  At 32 km we started rotating
the gondola with a period of 18 s to measure the CMB dipole. Data from
this calibration are shown in Fig 13.  The dipole measurement lasted
30 minutes which included $\approx 100$ rotations.  The data are well
fit by a model that includes the CMB dipole, thermal emission from
galactic dust, and a fraction of the signal from the 410 GHz channel
in the same row of the array.  This component of the signal is
attributed to emission from the atmosphere.  This component decreased
with time during the dipole measurement as the gondola rose from 32 to
36 km in altitude.  It only makes a significant contribution to the
first quarter of the data.

\begin{figure}[t!] 
\centerline{\epsfig{file=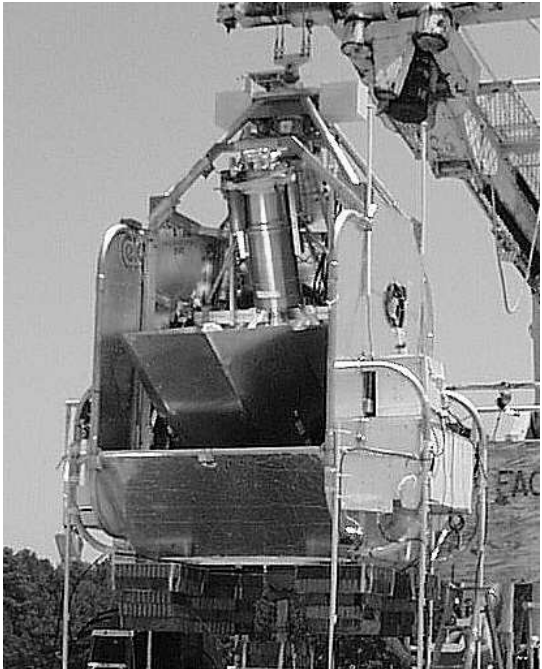, height=3.5in,angle=0}}
\caption{Photograph of the MAXIMA gondola shortly before launch.  The
receiver is visible in the center of photograph.  The scoop-shaped
aluminum ground shield can be seen, below the receiver.  The outside
of the gondola is covered with aluminum coated insulation attached to
the roll-bars.}
\vspace*{10pt}
\label{fig12}
\end{figure}

\begin{figure}[t!] 
\centerline{\epsfig{file=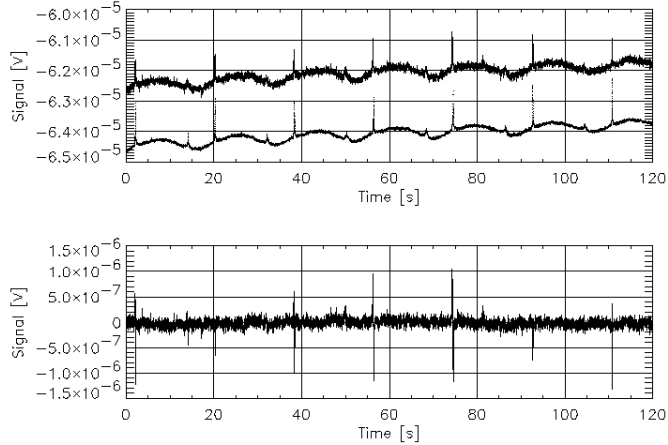, height=3in,angle=0}}
\caption{Observation of the CMB dipole during the MAXIMA-1 flight.
The top panel shows raw data from one 150 GHz channel at the top and a
model calculation at the bottom.  The model is a linear combination of
a dipole simulation, a galaxy simulation from an IRAS/DIRBE $100 ~\mu
m$ map\protect\cite{schlegel98}, data from the 410 GHz channel in the same row
in the array, slope, and a constant offset.  The signal correlated
with the 410 GHz channel may be due to atmospheric emission.  It
decreases with time during the observation as the gondola rises from
32 to 36 km, and only makes a significant contribution to the first
quarter of the data. The bottom panel shows the difference between the
model and the data.  The fit between model and data for this channel
give a reduced $\chi^2 \sim 1$. The galaxy crossings are not included
in the fit.}
\vspace*{10pt}
\label{fig13}
\end{figure}

After the CMB dipole measurement, we carried out two CMB anisotropy
scans which lasted a total of 3 hours.  The position of one beam
during this period is shown in Fig. 14.  We perform three temporal
modulations. First, the mirror scans $4^\circ$ p-p at 0.45 Hz.
Second, the gondola is scanned slowly in azimuth with a period of
$\approx 1$ minute with the elevation of the telescope held constant.
During the first and second segments the azimuth angle of the gondola
scanned $8^\circ$ and $5.6^\circ$ p-p respectively. Finally, the
telescope azimuth and elevation are changed roughly at the midpoint of
the observation to rescan the same region.  This second scan on the
sky occurs at an angle to the first due to sky rotation producing a
cross-linked pattern as shown in Fig. 14. This cross-linking of pixels
allows a two dimensional map of the sky to be reconstructed from one
dimensional scans in the presence of inevitable low-frequency noise in
the detectors.  The two-dimensional information that can be obtained
from such a map gives a more sensitive measurement of the power
spectrum than can be obtained from one-dimensional
information.\cite{theory} The total area covered by the observation is
122 deg$^2$ (0.3\% of the sky).  This region is one of the lowest dust
contrast regions of the sky based on calculations using the IRAS/DIRBE
$100 ~\mu m$ map of Schlegel {\it et al.}\cite{schlegel98}.

\begin{figure}[t!] 
\centerline{\epsfig{file=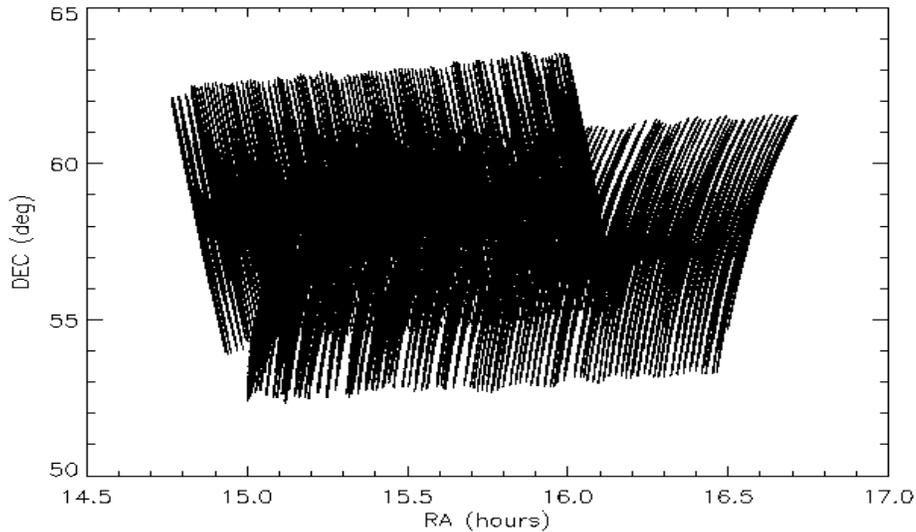,height=3in,width=5in,angle=0}}
\caption{Scan pattern for MAXIMA-1 flight.  Points represent the
position of the beam for one channel.  The total area covered by the
observation is 122 deg$^2$ (0.3\% of the sky).  The duration of the
observation was 3 hours.  This region is one of the lowest dust
contrast regions of the sky based on calculations using the IRAS/DIRBE
$100 ~\mu m$ map of Schlegel {\it et al.}\protect\cite{schlegel98}. Observation
strategy includes three temporal modulations. The mirror scans
$4^\circ$ p-p at 0.45 Hz, the gondola is slowly sawtooth scanned in
azimuth with a period of $\approx1$ minute with the elevation of the
telescope held constant, and the telescope elevation is changed
roughly at the midpoint of the observation to rescan the same region.}
\vspace*{10pt}
\label{fig14}
\end{figure}

During the last part of the flight we observed Jupiter to measure our
beam patterns and obtain an independent calibration.  We held the elevation
of the telescope constant while Jupiter slowly moved through the
elevation of the beams due to sky rotation.  The mirror modulation
provided azimuthal scans of Jupiter at many elevations.  One such scan
is shown in Fig. 7.  We made several hundred such scans for each
channel.  It should be possible to form a two-dimensional map of the
beam from these scans to accurately characterize the beam shapes.

The experiment was recovered after flight with little damage and is
being prepared for up to two flights from Palestine during the summer
of 1999.  Data analysis from MAXIMA-1 is in progress and results will
be presented in future publications.

\section*{acknowledgements}

We are grateful to R. Plambeck, R.F. Silverberg, and P. Timbie for
supplying the RF sources used for the sidelobe measurements.  We thank
S. Church, D. Finkbeiner, E. Hivon, D. Huguenin, and S. Meyer for
useful discussions.  D. Horsley and J. Wu made important contributions
to the attitude control system.  We are very grateful to the National
Scientific Balloon Facility in Palestine, Texas for their support
during the MAXIMA-0 and MAXIMA-1 campaigns.  This work was funded in
part by NSF cooperative agreement AST-9120005 and NASA grant
NAG5-4454.

\end{document}